\documentclass{nature}
\usepackage{amsmath,amssymb,graphicx,color,bm,soul}
\usepackage{textcomp} 
\usepackage{hyperref}
\makeatletter
\let\saved@includegraphics\includegraphics
\AtBeginDocument{\let\includegraphics\saved@includegraphics}
\renewenvironment*{figure}{\@float{figure}}{\end@float}
\makeatother

\title{Room temperature Organic Exciton-Polariton \\ Condensate in a Lattice}

\author{M. Dusel$^{1}\dagger^{*}$, S. Betzold$^{1}\dagger$, O. A. Egorov$^{2}$, S. Klembt$^{1}$, J. Ohmer$^{3}$, U. Fischer$^{3}$, S. H\"ofling$^{1,4}$ \& C. Schneider$^{1,*}$}

\begin{document}

\maketitle

\begin{affiliations}
 \item Technische Physik, Universit\"at W\"urzburg, Am Hubland, D-97074 W\"urzburg, Germany
 \item Institute of Condensed Matter Theory and Solid State Optics, Abbe Center of Photonics, Friedrich-Schiller-Universit\"at Jena, Max-Wien-Platz 1, D-07743 Jena, Germany
 \item Department of Biochemistry, Universit\"at W\"urzburg, Am Hubland, D-97074 W\"urzburg, Germany
 \item SUPA, School of Physics and Astronomy, University of St Andrews, St Andrews KY16 9SS, United Kingdom
 \item[$\dagger$ These authors contributed equally to this work] 
 \item[$*$ Corresponding author. Email: Christian.Schneider@uni-wuerzburg.de (C.~S.),]
 \item[Marco.Dusel@uni-wuerzburg.de (M.~D.)]
\end{affiliations}

\begin{abstract}
{Interacting Bosons, loaded in artificial lattices, have emerged as a modern platform to explore  collective manybody phenomena, quantum phase transitions and exotic phases of matter as well as to enable advanced on chip simulators. Such experiments strongly rely on well-defined shaping the potential landscape of the Bosons, respectively Bosonic quasi-particles, and have been restricted to cryogenic, or even ultra-cold temperatures. On chip, the GaAs-based exciton-polariton platform emerged as a promising system to implement and study bosonic non-linear systems in lattices, yet demanding cryogenic temperatures. In our work, we discuss the first experiment conducted on a polaritonic lattice at ambient conditions: We utilize fluorescent proteins as an excitonic gain material, providing ultra-stable Frenkel excitons. We directly take advantage of their soft nature by mechanically shaping them in the photonic one-dimensional lattice. We demonstrate controlled loading of the condensate in distinct orbital lattice modes of different symmetries, and finally explore, as an illustrative example, the formation of a gap solitonic mode, driven by the interplay of effective interaction and negative effective mass in our lattice. The observed phenomena in our open dissipative system are comprehensively scrutinized by a nonequilibrium model of polariton condensation. We believe, that this work is establishing the organic polariton platform as a serious contender to the well-established GaAs platform for a wide range of applications relying on coherent Bosons in lattices, given its unprecedented flexibility, cost effectiveness and operation temperature.}
\end{abstract}

The implementation of well-controllable lattice potentials for interacting quantum particles and Bosonic condensates\cite{ImBloch2008} is an engineering task of importance towards the realization of advanced classical\cite{Ozawa2019} and quantum simulators\cite{Yamamoto2010}. The possibilities to engineer Hamiltonians in well-defined experimental settings has furthermore served as an inspiration to explore topology in synthetic systems\cite{Stringari1996}. While ultra-cold atoms in optical lattices\cite{Nasci2012}, trapped ions\cite{Roos2012}, superconducting circuits\cite{Koch2012,Nori2011} and photonic on-chip architectures\cite{Laing2018} are considered as the leading platforms for a controlled experimental implementation and manipulation, they intrinsically suffer from a variety of serious drawbacks: The atomic approach relies on ultra-low temperatures, purely photonic approaches suffer from very small non-linearities, and superconducting circuits compose a serious technological challenge.

A hybrid approach, seeking to merge the best of all worlds, involves the implementation of strongly interacting photons on chip, in particular in the form of exciton-polaritons. Such hybrid excitation can inherit the low-loss nature of photons, and still acquire a notable non-linearity arising from the excitonic part\cite{Volz2019,Schade2019}.
Since the initial demonstration of Bose-Einstein condensation of microcavity exciton polaritons\cite{Kasrpzak2006}, one particular focus was set on engineering the potential landscape of polaritons, in the spirit of on-chip quantum simulation\cite{Schneider-ROPP16}. While cavity polaritons and their condensates have been observed in a variety of systems, including II-VI\cite{Kasrpzak2006}, III-V semiconductors\cite{BaliliScience}, layered materials\cite{Liu2014,Waldhorn2018}, as well as organic materials\cite{Plumhof2013,Daskalakis2014,Dietrich2016} due to the technological challenges required to precisely control their energy landscape, the vast majority of approaches towards periodic arrangements were conducted on the mature GaAs platform\cite{Schneider-ROPP16} .  
The degree of energy, position, coupling and phase control has now reached such a level, that enabled the first implementation of ultra-fast simulators of the X-Y Hamiltonian\cite{Berloff2017}, as well as synthetic topological Chern insulators\cite{Klembt2018} and topological lasers\cite{StJean2017}. Even electrical injection has recently accomplished\cite{Suchomel2018}. However, these experiments still require cooling by liquid helium, and (in most cases) highly advanced nano-technology for chip processing.

In this work, we address these two fundamental complications: We utilize fluorescent proteins as active material, hosting ultra-stable Frenkel excitons. Furthermore, we directly take advantage of the soft nature of the material, and mechanical shape it into the photonic lattice environment, which has the form of a one-dimensional lattice of tightly coupled photonic hemispheric cavities.  
We make the following, striking observations: 
The high quality of our device allows us to generate a close-to ideal bandstructure of room-temperature exciton-polaritons, dictated by the photonic lattice in the tight-binding regime. Bosonic condensation is facilitated at elevated pump densities, and by shaping the pump spot, we can load the condensate into distinct lattice modes. The subtle interplay between the repulsive polaritonic interaction, and the negative polariton mass yields the formation of a one-dimensional gap solitonic state, which we clearly identify in the realspace expansion of the condensate.

We further develop a broad understanding of these nonequilibrium effects in the framework of a numerical model based on solving the Gross-Pitaevskii equation in presence of gain, loss and noise.

\begin{figure}
	\centering
	\includegraphics[width=\textwidth]{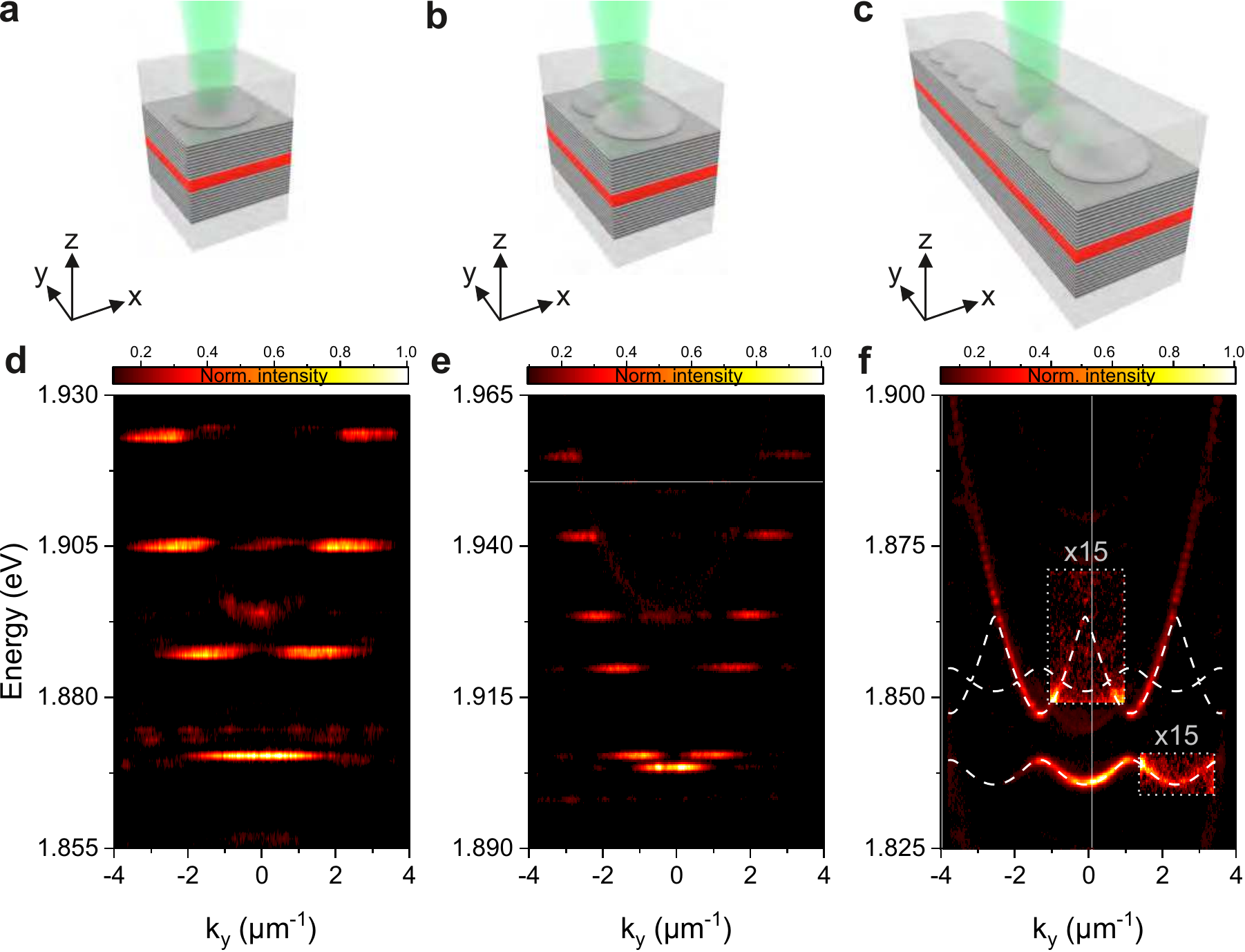}
	\caption{\textbf{Schematic images and angle-resolved measurements of our device.} \textbf{a - c,} Artistic illustration of single trap (\textbf{a}), molecular configuration (\textbf{b}) and  one-dimensional lattice (\textbf{c}).
		The hemispheric indentations are filled with the "mCherry" proteins (red). \textbf{d - f,} Angle-resolved photoluminescence spectrum of a single trapping site (\textbf{d}), as well as a  molecular configuration (\textbf{e}) and the one-dimensional lattice (\textbf{f}), proving the formation of a bandgap polariton spectrum resulting from evanescent coupling between the sites. White lines show calculated single-particle energy bands in the effective periodic potential of the depth -\,160\,meV. The measurements are recorded at a detuning between the cavity photon and exciton energy of $\Delta=E_C-E_X=$\,-\,100 meV (at $k=0$ of the ground Bloch-band). 
	\label{fig1}}
\end{figure}

Our studied device is composed of two distributed Bragg reflectors (DBRs), which sandwich an optical spacer layer filled with a thin film of the fluorescent protein "mCherry". The layer of mCherry is coated on the hemispheric micro-lenses, and the planar back-mirror mechanically presses the proteins into the indentations, prior to thermally evaporating the solvents.  An artistic sketch of the device is shown in Fig. \ref{fig1}a-c.  The DBRs each have 10.5 alternating pairs of SiO$_2$- and TiO$_2$-layers with a reflectivity of 99.9\,\% between 1.90\,eV (653\,nm) and 2.13\,eV (582\,nm). This configuration yields calculated mode quality factors exceeding 20000 and support strong coupling conditions with a Rabi-splitting of 240\,meV, as we analyze in the supplementary information S1. Prior to coating the top DBR, we prepared a plateau like area on the glass substrate with depth and diameter of about 500\,\textmu m and 4000\,\textmu m, respectively, by chemical wet etching. Then, lens-shaped indentations were defined by using ion beam milling. These hemispheric dimples have diameters ranging from 3\,\textmu m to 5\,\textmu m and depths between 100\,nm and 350\,nm. The shape of the micro-lenses confines the optical field to a spatial area with effective radii of 1.0\,\textmu m to 1.5\,\textmu m, depending on the radius of curvature of the lens. Consequently, the dispersion relation of cavity photons, and likewise, exciton-polaritons in such a trap acquires the canonical dispersion relation of a massive particle in a harmonic trap (see Fig. \ref{fig1}d), featuring a ladder of dispersion-less, discrete resonances with an approximately equal energy spacing. Recently, we have discussed that the trapping of exciton-polaritons in such a structure promotes the formation of room-temperature polariton condensation with high coherence\cite{Betzold2019}. However, the controllable confinement of polaritonic quasi-particles in a single trap also composes a key building block towards the design of more advanced potential landscapes. In the presence of a neighboring hemispheric trap (see Fig. \ref{fig1}e), the optical modes transform into the frequency-split molecular resonances, with a well-controllable molecular coupling\cite{Dufferwiel2015,Urbonas2016}. Next, we study a one-dimensional  periodic arrangement of potential traps under non-resonant excitation (532\,nm) in the low-density limit.

Fig. \ref{fig1}f  shows the far field emission spectra of such an lattice (trap size 5\,\textmu m, next neighbor distance 2.5\,\textmu m). The observed emission {\em below the condensation threshold} strongly differs from an isolated hemispheric lens, where the trapped modes are discrete in energy and momentum\cite{Betzold2019}. Evanescent coupling of the modes confined in the polariton traps leads to formation of a distinct bandgap spectrum due to the spatial periodicity of the structure. Up to two Bloch-bands are visible in the spectrum of the lower polariton, as shown in Fig. \ref{fig1}f. Because of the deep confinement, a gap between the ground and the first excited Bloch-band evolves as large as 5\,meV. The experimentally determined bandgap structure of the spectrum can be fitted by a spectrum of single-particle eigenstates in a periodic one-dimensional lattice following the methodology introduced in\cite{Winkler2015} (details on the model can be found in the methods section, and as well as supplementary information S2).

\begin{figure}
	\centering
	\includegraphics[width=\textwidth]{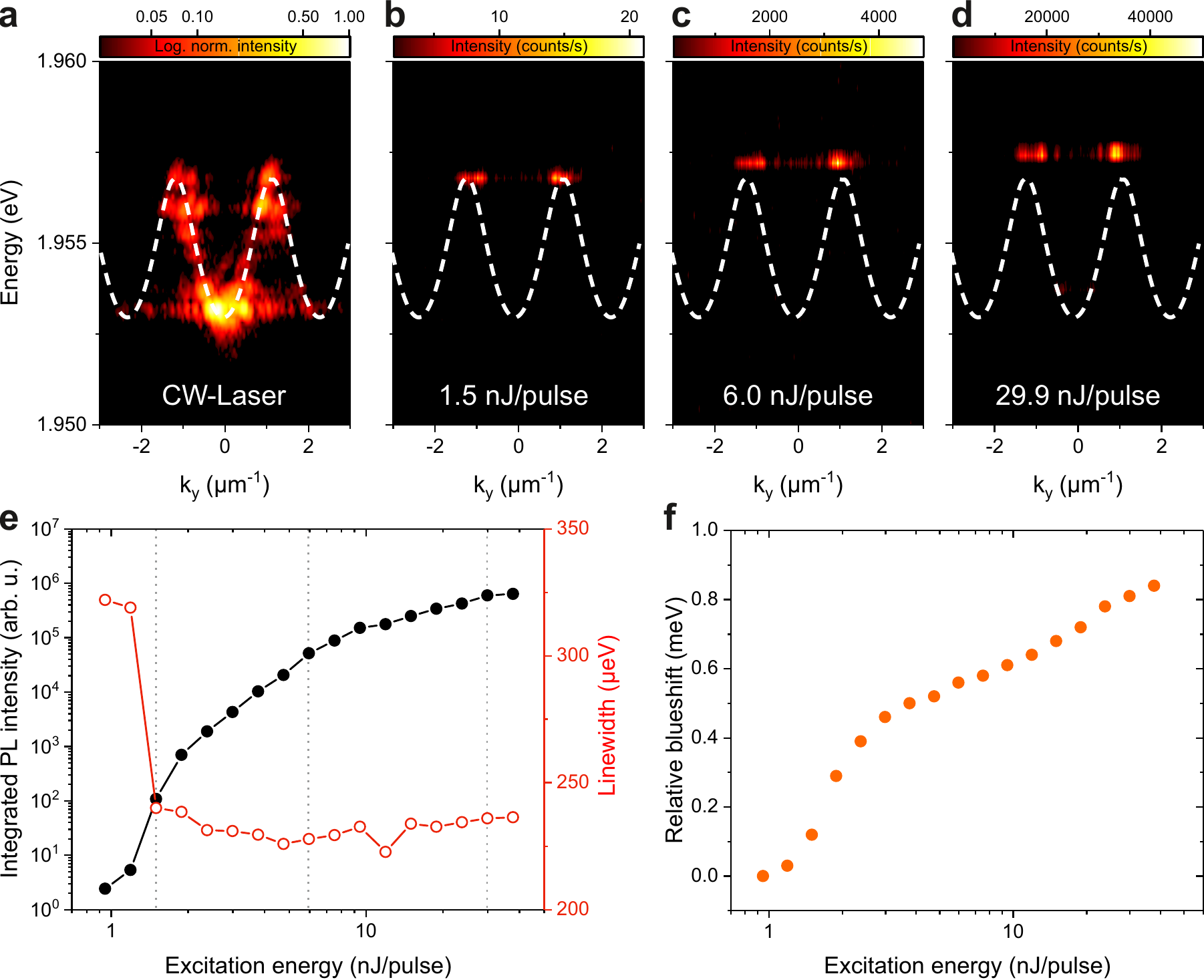}
	\caption{\textbf{Excitation power-dependent analysis of an one-dimensional lattice.} \textbf{a - d,} Far-field photoluminescence spectra recorded at various pump powers. Excitation with a continuous wave laser (\textbf{a}), the white line shows calculated single-particle energy s-band in the effective periodic potential. Angled-resolved spectra for pump powers at the condensation threshold (\textbf{b}, P\,=\,1.5\,nJ/pulse), above the threshold (\textbf{c}, P\,=\,6.0\,nJ/pulse) and far past the threshold (\textbf{d}, P\,=\,29.9\,nJ/pulse). The condensed mode moves into the gap. \textbf{e,} Integrated emission intensity (black) and linewidth (red) versus excitation energy. At P\,=\,1.5 nJ/pulse the linwidth drops to the resolution limit of the spectrometer whereas the intensity clearly features a superlinear increase. \textbf{f,} Excitation power-dependent blueshift of the mode. \label{fig2}}
	
\end{figure}

The high stability of the Frenkel Excitons makes the non-linear regime of bosonic condensation accessible. In the following experiment, we excite the one-dimensional lattice with a Gaussian spot (diameter $\approx$\,3\,\textmu m), focused on a hemispheric trap in the lattice. By increasing the pump power from P\,=\,0.95\,nJ/pulse over P\,=\,1.5\,nJ/pulse (Fig. \ref{fig2}b), P\,=\,6.0\,nJ/pulse (Fig. \ref{fig2}c) to P\,=\,29.9\,nJ/pulse (Fig. \ref{fig2}d), we experience a significant qualitative modification of the lattice spectrum. First, one can capture a significant, non-linear increase in the emission intensity, centered around the spectral region on top of the Bloch-band formed from photonic s-type wavefunctions. The non-linear threshold, characterized by the typical s-shape of the extracted peak area versus pump energy (see Fig \ref{fig2}e), the strong drop in the linewidth down to the resolution limit of our spectrometer, as well as the persisting blueshift of the mode above threshold puts our system well in the framework of room-temperature bosonic condensation. 

Once the condensation threshold is reached at P\,=\,1.5\,nJ/pulse (see Fig. \ref{fig2}b), in qualitative agreement with experiments conducted at 10\,K in significantly more shallow potentials\cite{Lai2007}, as well as in the tight-trapping limit\cite{Gao2018,Winkler2016}, the polariton condensate emerges at the vicinity of the BZ edges $k_{||}=\pm \pi/a$ and progressively shifts into the gap of the structure, with an approximate energy offset of $\approx$\,0.56\,meV with respect to the maximum of the ground band (Fig.~\ref{fig2}c). With increasing power of the pump laser, the energy of the condensate shifts further into the gap (Fig. ~\ref{fig2}d), while remaining localized at the edges of the BZ in the momentum space.

\begin{figure}
	\centering
	\includegraphics[width=\textwidth]{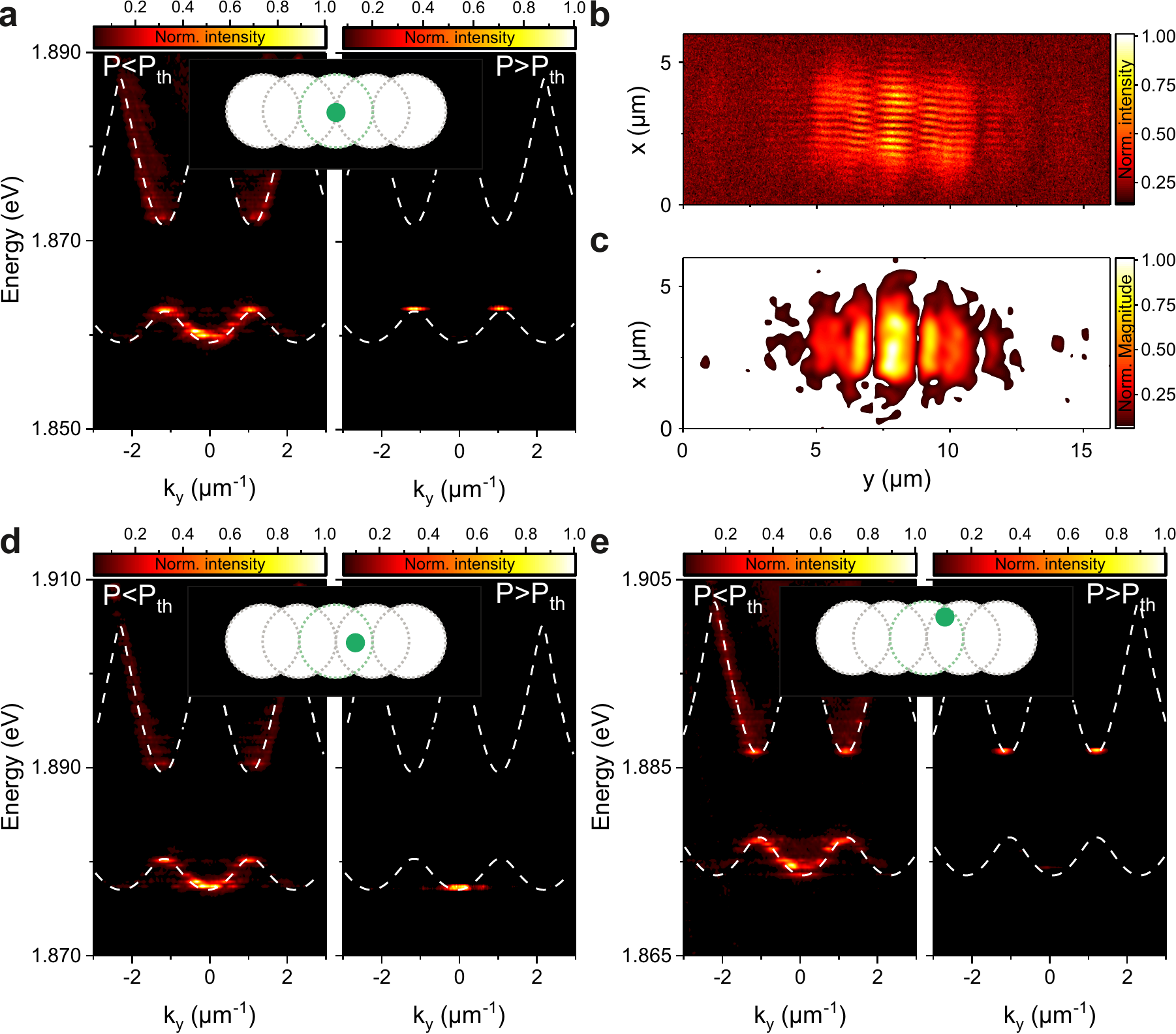}
	\caption{\textbf{Controlled loading of the polariton condensate in the lattice. {\em off-site} versus {\em on-site} pumping and formation of  corresponding gap state condensates.} \textbf{a, d, e,} Far-field spectra below (left) and above (right) the condensation threshold, corresponding with the pump-potential alignment inserted with a sketch of the position of the pump laser spot (green) aligned with respect to the polariton potential. \textbf{b,} Spatial coherence measurement under {\em on-site} excitation above threshold and \textbf{c,} extracted coherent magnitude of the interference pattern. \label{fig3}}
	
\end{figure}

Polaritonic condensation in our lattice is not restricted to the anti-binding orbital s-mode of our lattice. It is well-controllable by tuning the overlap between the pump laser and the optical mode, making our system extraordinarily versatile: As shown in Fig. \ref{fig3}a,  similarly to the scenario discussed above, the {\em on-site} pumping condition loads the polaritonic condensate into the anti-binding s-mode, followed by a subsequent energy shift into the gap. 
The spatial coherence of this mode can be analyzed by measuring its correlation function $g^{1}(r,r^{'})$. We use a Michelson interferometer with a variable path length in the mirror-retroreflector configuration, which overlaps the realspace luminescence from the device with its mirror image on a beam-splitter, and combines them on a high resolution CCD camera\cite{Kasrpzak2006}. The resulting image (Fig. \ref{fig3}b) shows a distinct interference pattern, in particular around the area of the pump spot. The spatially resolved visibility of the interference fringes, plotted in Fig. \ref{fig3}c, reflects that the coherent mode expands over various lattice sites.
In strong contrast, {\em off-site} pumping conditions result in a significantly different phenomenological behavior at any moderate pump power above the threshold of condensation. In this setting, the condensate occurs in the binding ground-state of the lattice (see Fig. \ref{fig3}d). 
Again, by laterally displacing the pump-laser beam, yielding an improved overlap with the p-type orbital mode of the second dispersive band, the condensation process can even be triggered on top of the bandgap (see Fig. \ref{fig3}e). 

The observed behavior of the condensed state in Fig. \ref{fig2} and Fig. \ref{fig3}b ({\em on-site} pumping) hints at the presence of a spatially localized polariton gap state, which has previously been reported in\cite{Tanese2013, Gao2018,Winkler2016} for GaAs-based exciton-polariton condensates in photonic waveguide arrays\cite{Kivshar2003} and atomic Bose-Einstein condensates in optical lattices\cite{Oberthaler2004,Louis2003,Matuszewski2006}.

\begin{figure}
\centering
\includegraphics[width=\textwidth]{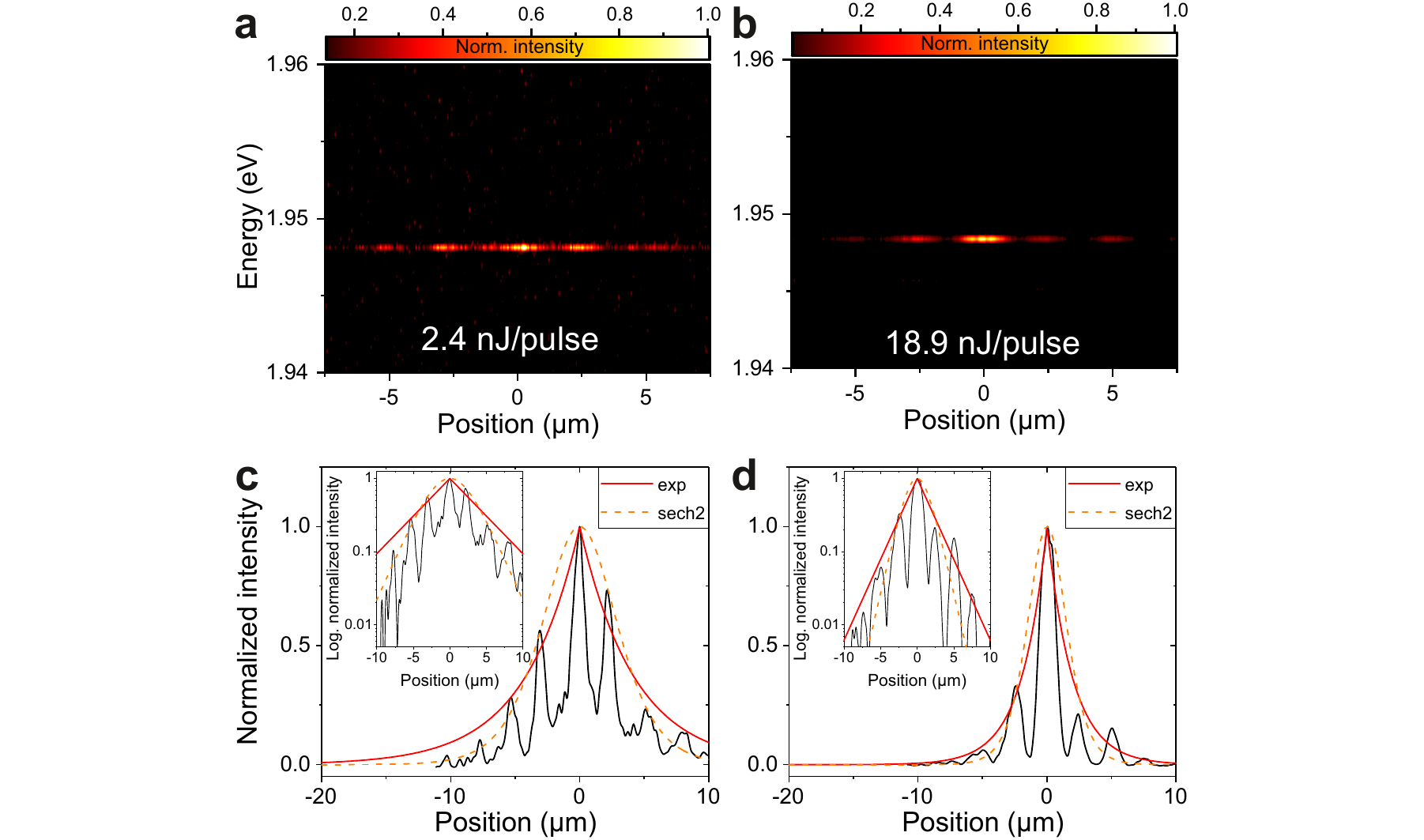}
\caption{\textbf{Self localization of a gap-solitonic mode.}  \textbf{a, b,} Energy-resolved realspace images under {\em on-site} pumping conditions little above threshold (\textbf{a}, P\,=\,2.4\,nJ/pulse), and far above threshold (\textbf{b}, P\,=\,18.9\,nJ/pulse). \textbf{c, d,} The intensity traces are fitted by a $sech^{2}$ and an exponential function. For the lower excitation power the $sech^{2}$-function fits to the intensity profile while for higher excitation power the exponential function fits better. In the insets, the same profiles are shown in log scale. Spatial narrowing of the extension of the condensate characterize the gap-soliton mode.  
\label{fig4}}

\end{figure}

Solitonic behavior is expected to manifest in the localization- and expansion of the condensate in our lattice in the high-density regime. 
Here, we studied the luminescence from our device under {\em on-site} pumping conditions and projected its real-space distribution onto a high-resolution camera. The leakage from the microcavity yields a stochastic decay of the quasiparticles, and results in a characteristic decrease of the luminescence intensity  away from the pump spot. Fig. \ref{fig4}a and Fig. \ref{fig4}b depict the energy-resolved realspace distribution of the emitted photoluminescence. One can clearly identify the luminescence maxima from the single sites in the lattice (see also angle-resolved measurements in supplementary information S3).  While the pump power in Fig. \ref{fig4}a was chosen slightly above the onset of polaritonic condensation (P\,=\,2.4\,nJ/pulse), in Fig. \ref{fig4}b the condensation threshold was exceeded by approximately one order of magnitude (P\,=\,18.9\,nJ/pulse). The most dramatic effect of the condensation into the gap state can be seen from the intensity traces in Fig. \ref{fig4}c and Fig. \ref{fig4}d. As the excitation power increases by one order of magnitude, the spatial extension of the condensate becomes substantially narrower. This effect of self-localization, leading to an effective reduction of the condensate expansion, is canonical for the formation of a gap solitonic mode and is an excellent agreement with our numeric model (see supplementary information S2). We furthermore witness a change in the shape of the condensed mode which transits from a squared hyperbolic secant profile (indicating the role of interactions in our systems) at moderate pump power to a dominantly exponential decay, in striking similarity to earlier experiments conducted on GaAs-based cavity polaritons in modulated wire cavities\cite{Tanese2013}.

A remarkable feature in our system, which was pointed out in an earlier work\cite{Betzold2019}, is the nature of the interaction in the fluorescent protein system, which is crucial for the formation of a gap-solitonic state\cite{Ostrovskaya2013}. While inter-particle repulsion in GaAs-based cavities is mostly induced by polariton-polariton interaction and interaction with the exciton reservoir, the density dependent blueshift in Frenkel Excitons mostly arises from the screening of the Rabi-splitting with elevated density\cite{Betzold2019}. As our experiments confirm, the nature of interaction is not of relevance for the observed phenomenological behavior for our system in the nonlinear regime.

\section*{Conclusions} We have established the organic platform of exciton-polaritons as a new, attractive player to study bosons, and in particular non-linear bosonic condensates in optical lattices. Our platform is low-cost, flexible and can be operated at ambient conditions. Here, we already observe the full bandstructure of polaritonic Bloch-bands emerging in the regime of tight binding, the controlled loading of condensates into Bloch-bands with specific orbital symmetry, and the excitation of localized gap states driven by the intrinsic non-linearity of our system.  
Our work can be extended straight forwardly to arbitrary two-dimensional lattice geometries, as well as synthetic time dimensions, for on-chip bosonic annealing, quantum simulation and topological polaritonics.

\begin{methods}
\subsection{Experimental Setup}
Our optical setup can be used concurrent in near-field (spatially) and far-field (momentum space) resolved spectroscopy and imaging. We used two lasers to excite our sample. First a 532\,nm continuous-wave diode laser and second  a wavelength-tunable optical parametric oscillator system with ns-pulses tuned to 532\,nm that is resonant with the first Bragg minimum of the top mirror of our sample. 
For angle resolved measurements we used a Fourier imaging configuration. Here a Fourier lens collecting the angle-dependent information in the back-focal plane of the microscope objective (50x, 0.42\,NA). To obtain a spatially resolved signal we used the Fourier configuration with an extra lens. The emission is filtered by a 550\,nm longpass filter. Both the near-field and the far-field create an image in the focus plane of the spectrometer system. The slit in entrance plane is closed to exclude a narrow section of the image. The emitted signal is dispersed by an optical grating with 1200\,lines/mm with a spectral resolution of about 230\,\textmu m for energies around 2\,eV. 
For the Michelson measurements the emitted signal is deflected through a beam splitter and overlaid itself with a retroreflector and a mirror.

\subsection{Numerical Modelling}
For numerical calculations of the polariton dynamics in organic lattices we consider a two-dimensional mean-field model, which has been widely used for simulation of inorganic semiconductor cavities\cite{RefWoutersPRL991404022007, Wouters2008} and very recently for organic systems\cite{Bobrovska2018}. This model consists of the open-dissipative Gross-Pitaevskii (GP) equation for the condensate wave-function incorporating stochastic fluctuations and coupled to the rate equation for the excitonic reservoir created by the off-resonant pump\cite{Wouters2008, Winkler2016}.

In particular our theoretical analysis confirms that the gap state bifurcates from the upper edge of the band characterised by the "staggered" phase, and therefore inherits the characteristic $\pi$-phase shift between neighbouring density peaks\cite{Ostrovskaya2013}. For the further details on numerical modelling we refer to the supplementary informations. 

\end{methods}

\pagenumbering{gobble}

\end{document}